\documentclass[sigconf]{acmart}

\usepackage{booktabs,multirow,array} 
\usepackage[ruled,linesnumbered]{algorithm2e}
\usepackage[english]{babel}

\usepackage{enumitem}
\usepackage{booktabs} 
\usepackage{multirow}
\usepackage{array}
\usepackage{subcaption}
\usepackage{adjustbox}
\usepackage{graphicx}
\usepackage[justification=justified,skip=5pt]{caption}

\usepackage{amsmath}
\usepackage{amsfonts}
\usepackage{bbm}
\usepackage{bm}
\usepackage{mathtools}
\usepackage{scalerel} 
\usepackage[ruled,linesnumbered]{algorithm2e}

\usepackage{soul}
\setstcolor{red}
\usepackage[english]{babel}
\usepackage{lipsum}
\usepackage{enumitem}
\usepackage[justification=justified, skip=0pt]{caption}

\makeatletter
\g@addto@macro\normalsize{%
  \abovedisplayskip 5pt plus 2pt minus 3pt%
  \belowdisplayskip \abovedisplayskip
  \abovedisplayshortskip 5pt plus2pt  minus3pt%
  \belowdisplayshortskip 5pt plus2pt minus3pt%
}

\makeatother

\newcommand\blfootnote[1]{%
\begingroup 
\renewcommand\thefootnote{}\footnote{#1}%
\addtocounter{footnote}{-1}%
\endgroup 
}
\DeclareMathOperator*{\argmax}{arg\,max}
\makeatletter
\def\BState{\State\hskip-\ALG@thistlm}
\makeatother

\copyrightyear{2021}
\acmYear{2021}
\setcopyright{acmcopyright}\acmConference[SIGIR '21]{Proceedings of the 44th International ACM SIGIR Conference on Research and Development in Information Retrieval}{July 11--15, 2021}{Virtual Event, Canada}
\acmBooktitle{Proceedings of the 44th International ACM SIGIR Conference on Research and Development in Information Retrieval (SIGIR '21), July 11--15, 2021, Virtual Event, Canada}
\acmPrice{15.00}
\acmDOI{10.1145/3404835.3462913}
\acmISBN{978-1-4503-8037-9/21/07}

\begin{document}

\fancyhead{}

\title{Unified Conversational Recommendation Policy Learning via Graph-based Reinforcement Learning}

\author{Yang Deng$^{1}$, Yaliang Li$^2$, Fei Sun$^2$, Bolin Ding$^2$, Wai Lam$^1$}
\affiliation{%
  \institution{\textsuperscript{\rm 1}The Chinese University of Hong Kong, \textsuperscript{\rm 2}Alibaba Group}
\institution{\{ydeng, wlam\}@se.cuhk.edu.hk,
 \{yaliang.li, ofey.sf, bolin.ding\}@alibaba-inc.com}}

\renewcommand{\authors}{Yang Deng, Yaliang Li, Fei Sun, Bolin Ding, Wai Lam}

\begin{abstract}
Conversational recommender systems (CRS) enable the traditional recommender systems to explicitly acquire user preferences towards items and attributes through interactive conversations. 
Reinforcement learning (RL) is widely adopted to learn conversational recommendation policies to decide what attributes to ask, which items to recommend, and when to ask or recommend, at each conversation turn. 
However, existing methods mainly target at solving one or two of these three decision-making problems in CRS with separated conversation and recommendation components, which restrict the scalability and generality of CRS and fall short of preserving a stable training procedure. 
In the light of these challenges, we propose to formulate these three decision-making problems in CRS as a unified policy learning task. 
In order to systematically integrate conversation and recommendation components, we develop a dynamic weighted graph based RL method to learn a policy to select the action at each conversation turn, either asking an attribute or recommending items. 
Further, to deal with the sample efficiency issue, we propose two action selection strategies for reducing the candidate action space according to the preference and entropy information. 
Experimental results on two benchmark CRS datasets and a real-world E-Commerce application show that the proposed method not only significantly outperforms state-of-the-art methods but also enhances the scalability and stability of CRS. 
\end{abstract}

%
%



\maketitle

\begin{figure*}
\setlength{\abovecaptionskip}{2pt}   
\setlength{\belowcaptionskip}{2pt}
\centering
\includegraphics[width=\textwidth]{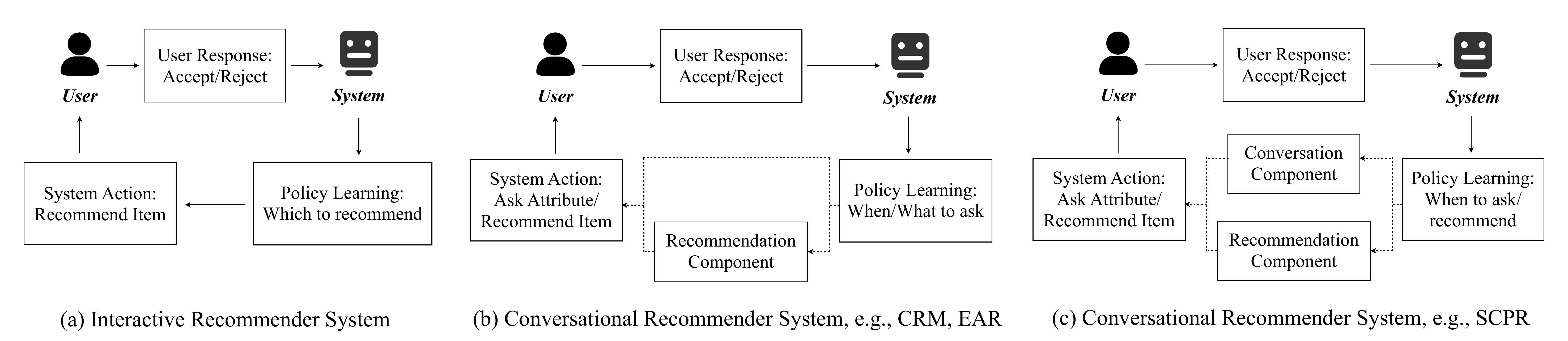}
\caption{Illustration of policy learning frameworks for IRS and CRS, including CRM~\cite{sigir18-crm}, EAR~\cite{wsdm20-ear}, and SCPR~\cite{kdd20-scpr}.}
\label{review}
\end{figure*}

\section{Introduction}
Conversational recommender systems (CRS) aim to learn user's preferences and make recommendations through interactive conversations~\cite{cikm18-saur,wsdm20-ear,nips18-redial}. 
Since it has the natural advantage of explicitly acquiring user's preferences and revealing the reasons behind recommendation, CRS has become one of the trending research topics for recommender systems and is gaining increasing attention. 
Unlike traditional recommender systems~\cite{icdm10-fm,www17-mf,recbole} or interactive recommender systems (IRS)~\cite{wsdm2020-irs-rl,sigir20-graph-irl}, which mainly focus on solving the problem of which items to recommend, there exists generally the other two core research questions for CRS~\cite{sigir20-tutorial}, namely what questions to ask and when to ask or recommend.
Recent works have demonstrated the importance of interactivity of asking clarifying questions in CRS~\cite{kdd16-convbandit,cikm18-saur,sigir20-qrs}.
More importantly, deciding when to ask or recommend is the key to coordinating conversation and recommendation for developing an effective CRS~\cite{sigir18-crm,wsdm20-ear,kdd20-scpr}.

Different problem settings of CRS have been proposed, either from the perspective of dialog systems, being a variation of task-oriented dialog~\cite{nips18-redial,coling2020-topic-crs,lei2018sequicity},  or from the perspective of recommender systems, being an enhanced interactive recommender system~\cite{sigir18-crm,kdd18-q&r,wsdm20-ear}. 
In this work, we study the multi-round conversational recommendation (MCR) setting~\cite{sigir18-crm,wsdm20-ear}, where the system asks questions about users' preferences on certain attributes or recommends items multiple times, and the goal is to make successful recommendation with the minimum number of interactions. 
\blfootnote{Work done when Yang Deng was an intern at Alibaba. This work was supported by Alibaba Group through Alibaba Research Intern Program, and a grant from the Research Grant Council of the Hong Kong Special Administrative Region, China (Project Codes: 14200719).}

In MCR scenario, the CRS is typically formulated as a multi-step decision-making process and solved by reinforcement learning (RL) methods for policy learning~\cite{sigir18-crm,wsdm20-ear,kdd20-scpr}. As shown in Figure~\ref{review}(a), RL-based IRS is only required to learn the policy to decide which items to recommend. However, the situation is more complicated in CRS, since there are two components that need to be coherently considered~\cite{arxiv20-crs}, i.e., conversation and recommendation components. Existing methods CRM~\cite{sigir18-crm} and EAR~\cite{wsdm20-ear} employ policy gradient~\cite{nips99-pg} to improve the strategies of when and what attributes to ask, while the recommendation decision is made by an external recommendation model. In order to reduce the action space in policy learning, another state-of-the-art method SCPR~\cite{kdd20-scpr} only considers learning the policy of when to ask or recommend, while two isolated components are responsible for the decision of what to ask and which to recommend. The policy learning frameworks of these two kinds of CRS are presented in Figure~\ref{review}(b) and \ref{review}(c). 

Despite the effectiveness of these methods, there are some issues remained to be tackled for real-world applications: 
(i) The models trained with existing CRS methods lack generality to different domains or applications, since there are three different decision-making processes to be considered in CRS, including what attributes to ask, which items to recommend, and when to ask or recommend. It requires extra efforts to train an offline recommendation model~\cite{kdd20-scpr,wsdm20-ear} or pretrain the policy network with synthetic dialogue history~\cite{wsdm20-ear,sigir18-crm}.
(ii) The policy learning is hard to converge, since the conversation and recommendation components are isolated and lack of mutual influence during the training procedure.

\begin{figure}
\setlength{\abovecaptionskip}{2pt}   
\setlength{\belowcaptionskip}{2pt}
\centering
\includegraphics[width=0.3\textwidth]{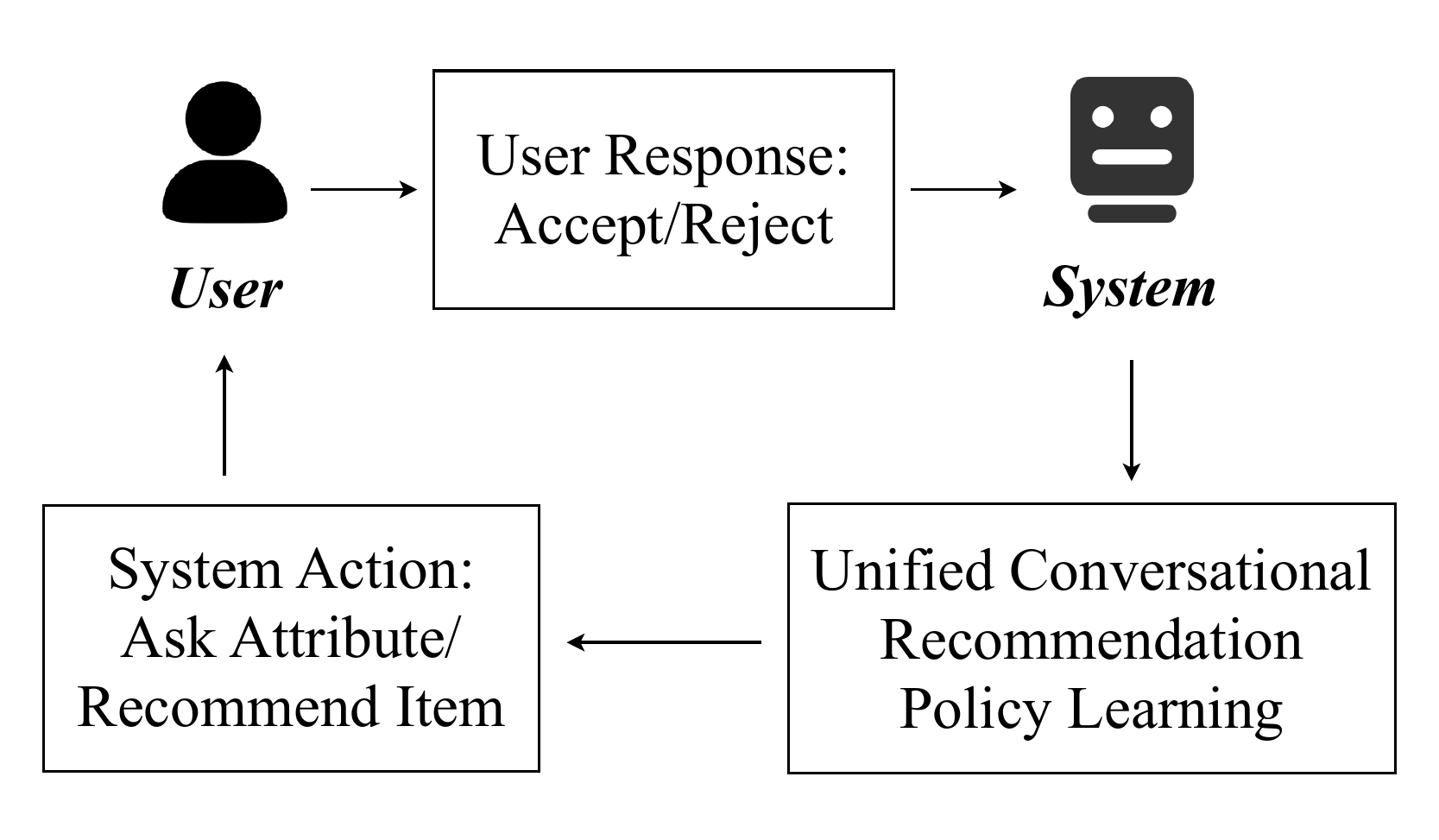}
\caption{Illustration of unified policy learning for CRS.}
\label{overview}
\vspace{-0.3cm}
\end{figure}

To address these issues, in this work, we formulate the aforementioned three separated decision-making processes in CRS as a unified policy learning problem to harness the ultimate goal of CRS as well as fill the gap between the recommendation and conversation components during the training procedure. 
Such unified conversational recommendation policy learning (UCRPL) aims at learning a unified policy to decide the action, either ask an attribute or recommend items, at each conversation turn to maximize the cumulative utility over the whole MCR process. The overview of the proposed unified policy learning for CRS is depicted in Figure~\ref{overview}.

However, the UCRPL problem comes along with two challenges: (i) How to systematically combine conversation and recommendation components for unified policy learning? (ii) How to deal with sample efficiency issues? As the action space becomes overwhelmingly large in UCRPL, including all available attributes and items, it requires tremendous amount of interaction data to learn the optimal policy. 
Fortunately, the graph structure, capturing the rich correlated information among different types of nodes (i.e., users, items, and attributes), enables us to discover collaborative user preferences towards attributes and items. 
Thus, we can leverage the graph structure to integrate the recommendation and conversation components as an organic whole, where the conversation session can be regarded as a sequence of nodes maintained in the graph to dynamically exploit the conversation history for predicting the action at the next turn. 
On the other hand, although the connectivity of the graph can also be used to eliminate invalid actions by path reasoning~\cite{kdd20-scpr}, there are still a large number of candidates left for action searching. Since users are not likely to be interested in all items and attributes, we can focus on the potentially important candidates to improve the sample efficiency of UCRPL.

To this end, we propose a novel and adaptive graph-based reinforcement learning framework, namely \textbf{UNI}fied \textbf{CO}nversational \textbf{R}ecomme\textbf{N}der (\textbf{UNICORN}). Specifically, due to the evolving nature of CRS, we leverage a dynamic weighted graph to model the changing interrelationships among users, items and attributes during the conversation, and consider a graph-based Markov Decision Process (MDP) environment for simultaneously handling the decision-making of recommendation and conversation. 
Then we integrate graph-enhanced representation learning and sequential conversation modeling to capture dynamic user preferences towards items and attributes.  
In addition, two simple yet effective action selection strategies are designed to handle the sample efficiency issue. Instead of enumerating the whole candidate item and attribute set, we adopt preference-based item selection and weighted entropy-based attribute selection strategies to only consider potentially important actions.

In summary, the contributions of this paper are as follows:
\begin{itemize}[leftmargin=*,topsep=4pt]
    \item We formulate three separated decision-making processes in conversational recommender systems, namely when to ask or recommend, what to ask, and which to recommend, as a unified conversational recommendation policy learning problem.
    \item To tackle the challenges in UCRPL problem, we propose a novel and adaptive reinforcement learning framework, namely \textbf{UNI}fied \textbf{CO}nversational \textbf{R}ecomme\textbf{N}der (\textbf{UNICORN}), based on dynamic weighted graph. To handle the sample efficiency issue, we further design two simple yet effective action selection strategies. 
    \item Experimental results show that the proposed method significantly outperforms state-of-the-art CRS methods across four public benchmark datasets and a real-world E-Commerce application. 
\end{itemize}


\section{Related Works}

\textbf{Conversational Recommendation.}
Based on the problem settings, current CRS studies can be categorized into four directions~\cite{sigir20-tutorial,convrec-survey}: 
(1) Exploration-Exploitation Trade-offs for
Cold-start Users~\cite{kdd16-convbandit,www20-convbandit,li2020seamlessly}. These approaches leverage bandit approaches to balance the exploration and exploitation trade-offs for cold-start users in conversational recommendation scenarios. 
(2) Question-driven Approaches~\cite{cikm18-saur,sigir20-qrs,kdd18-q&r} aim at asking questions to users to get more information about their preferences, which is often addressed as ``asking clarification/clarifying question''.  (3) Dialogue Understanding and Generation~\cite{kdd20-redial-kg,nips18-redial,coling2020-topic-crs,acl20-muulti-type-crs}. These studies focus on how to understand users' preferences and intentions from their utterances and generate fluent responses so as to deliver natural and effective dialogue actions.  (4) Multi-round Conversational Recommendation~\cite{sigir18-crm,wsdm20-ear,kdd20-scpr}. Under this problem setting, the system asks questions about the user’s preferences or makes recommendations multiple times, with the goal of  achieving engaging and successful recommendations with fewer turns of conversations. Among these problem settings, we focus on the MCR problem.

\noindent\textbf{RL in Recommendation.}
Reinforcement learning (RL) has been widely introduced into recommender systems due to its advantage of considering users' long-term feedbacks~\cite{www18-rl-newsrec,kdd18-rl-negativefeedback}.
RL-based recommendation formulates the recommendation procedure as an MDP of the interactions between the user and a recommendation agent, and employs RL algorithms to learn the optimal recommendation strategies~\cite{jmlr05,www18-rl-newsrec,kdd18-rl-negativefeedback,Pei:www19:Value}. Recent works on sequential recommendation~\cite{sigir20-rl-seqrec,sigir20-graph-rl-seqrec} and interactive recommendation~\cite{nips19-irs-rl,wsdm2020-irs-rl,sigir20-graph-irl} adopt RL to capture users' dynamic preferences for generating accurate recommendations over time. The goal of these approaches typically is to learn an effective policy for determining which items to recommend. As for CRS, RL-based methods are adopted to improve the strategies of the other two decision processes, including (i) what attributes to ask~\cite{sigir18-crm,wsdm20-ear} and (ii) when to ask or recommend~\cite{kdd20-scpr}. 
In order to  simplify the overall framework of MCR with better scalability and generality, we formulate these three core decision processes in CRS as a unified policy learning problem.

\noindent\textbf{Graph-based Recommendation.}
Graph-based recommendation studies mainly leverage the graph structure for two purposes. The first one is to enhance the recommendation performance by graph-based representation learning, including exploiting the structure information for collaborative filtering~\cite{sigir19-graph-cf,recsys18-graph-cf,sigir20-lightgcn}, and adopting knowledge graph embeddings as rich context information~\cite{kdd16-kg,sigir18-kg}. The other group of studies models recommendation as a path reasoning problem for building explainable recommender systems~\cite{www19-explain,aaai19-explain}. Recent years have witnessed many successful applications of graph-based RL methods on different scenarios of recommender systems~\cite{www20-graph-rl,sigir19-graph-rl,sigir20-graph-gcqn,sigir20-graph-irl,sigir20-graph-rl-demo,kdd20-scpr}. 
For example, \citet{sigir19-graph-rl} employ a policy-guided graph search method to sample reasoning paths for recommendation, which is enhanced with adversarial actor-critic for demonstration-guided path reasoning~\cite{sigir20-graph-rl-demo}. \citet{sigir20-graph-gcqn} and \citet{sigir20-graph-irl} employ graph convolutional network (GCN)~\cite{iclr17-gcn} for state representation learning to enhance the performance of traditional RL methods on recommendation policy learning. In this work, we study graph-based RL method for the conversational recommendation scenario based on a dynamic weighted graph.

\section{Problem Definition}
\textbf{Multi-round Conversational Recommendation.}
In this work, we focus on the multi-round conversational recommendation (MCR) scenario~\cite{wsdm20-ear,kdd20-scpr}, the most realistic conversational recommendation setting proposed so far, in which the CRS is able to ask questions about attributes or make recommendations multiple times. 

Specifically, on the system side, the CRS maintains a large set of items $\mathcal{V}$ to be recommended, and each item $v$ is associated with a set of attributes $\mathcal{P}_v$. In each episode, a conversation session is initialized by a user $u$ specifying an attribute $p_0$. Then, the CRS is free to ask the user's preference on an attribute selected from the candidate attribute set $\mathcal{P}_{\mathrm{cand}}$ or recommend a certain number of items (e.g., top-$K$) from the candidate item set $\mathcal{V}_{\mathrm{cand}}$. 
Following the assumptions from \citet{wsdm20-ear}, the user preserves clear preferences towards all the attributes and items. Thus, the user will respond accordingly, either accepting or rejecting the asked attributes or the recommended items. The CRS updates the candidate attribute and item sets, and decides the next action based on the user response. 
The system-ask and user-respond process repeats until the CRS hits the target item or reaches the maximum number of turn $T$.

\noindent\textbf{Unified Conversational Recommendation Policy Learning.}
Multi-round conversational recommendation aims to make successful recommendations within a multi-round conversation session with the user. At each timestep $t$, according to the observation on past interactions, the CRS selects an action $a_t$, either asking an attribute or recommending items. In return, the user expresses his/her feedback (accept or reject). This process repeats until the CRS hits the user-preferred items or reaches the maximum number of turn $T$. Such MCR task can be formulated as a Markov Decision Process (MDP). The goal of the CRS is to learn a policy $\pi$ maximizing the expected cumulative rewards over the observed MCR episodes as
\begin{equation}
    \pi^*=\argmax\nolimits_{\pi\in\Pi}\mathbb{E}\left[\sum\nolimits_{t=0}^Tr(s_t,a_t)\right],
\end{equation}
where $s_t$ is the state representation learned from the system state and the conversation history, $a_t$ is the action that the agent takes at timestep $t$, and $r(\cdot)$ is the intermediate reward, abbreviated as $r_t$.

\section{Methodology}\label{sec:method}
\begin{figure*}
\setlength{\abovecaptionskip}{2pt}   
\setlength{\belowcaptionskip}{2pt}
\centering
\includegraphics[width=\textwidth]{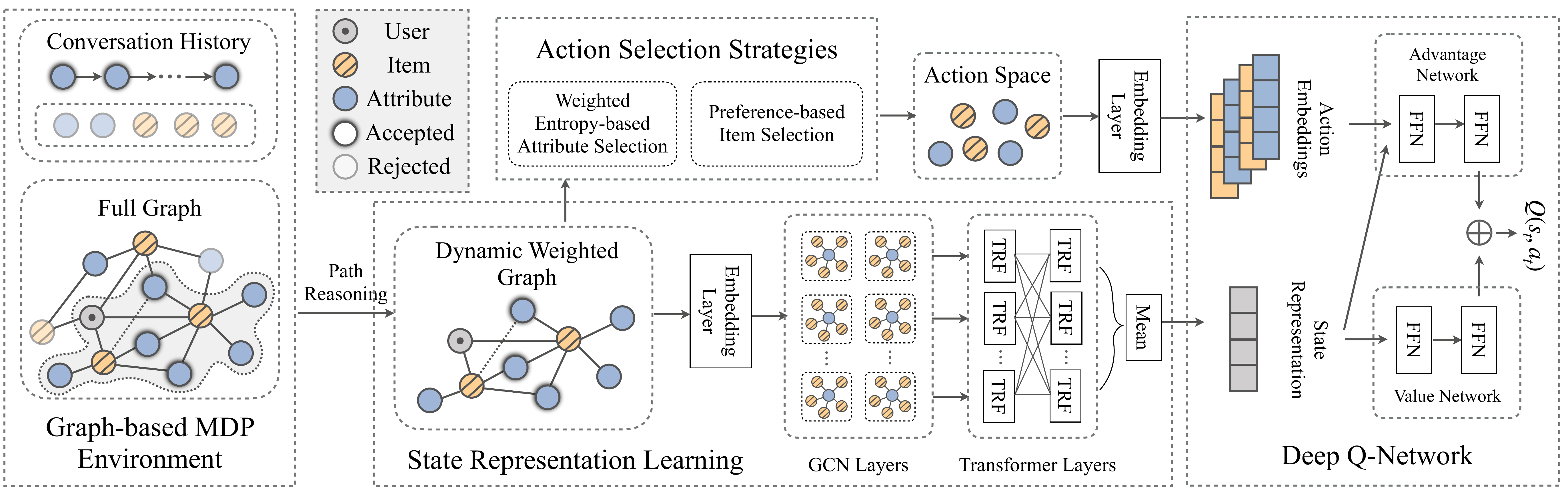}
\caption{Overview of the proposed method, UNICORN, for unified conversational recommendation policy learning. }
\label{method}
\end{figure*}

The overview of the proposed method, UNICORN, is depicted in Figure~\ref{method}, which consists of four main components: Graph-based MDP Environment, Graph-enhanced State Representation Learning, Action Selection Strategy, and Deep Q-Learning Network.

\subsection{Graph-based MDP Environment}
The MDP environment is responsible for informing the agent about the current state and possible actions to take, and then rewards the agent based on how the current policy fits the observed user interactions. Formally, the MDP environment can be defined by a tuple $(\mathcal{S},\mathcal{A},\mathcal{T},\mathcal{R})$, where $\mathcal{S}$ denotes the state space, $\mathcal{A}$ denotes the action space, $\mathcal{T}:\mathcal{S}\times\mathcal{A}\rightarrow \mathcal{S}$ refers to the state transition function, and $\mathcal{R}:\mathcal{S}\times\mathcal{A}\rightarrow \mathbb{R}$ is the reward function. 

\subsubsection{\textbf{State}}
As for the graph-based MDP environment, the state $s_t\in\mathcal{S}$ at timestep $t$ is supposed to contain all the given information for conversational recommendation, including the previous conversation history and the full graph $\mathcal{G}$ that includes all the users, items, and attributes. 
Given a user $u$, we consider two major elements:
\begin{equation}
    s_t = [\mathcal{H}_u^{(t)},\mathcal{G}_u^{(t)}],
\end{equation}
where $\mathcal{H}_u^{(t)} = [\mathcal{P}_u^{(t)}, \mathcal{P}_{\mathrm{rej}}^{(t)}, \mathcal{V}_{\mathrm{rej}}^{(t)}]$ denotes the conversation history until timestep $t$, and $\mathcal{G}_u^{(t)}$ denotes the dynamic subgraph of $\mathcal{G}$ for the user $u$ at  timestep $t$ (The graph construction will be introduced in Section~\ref{sec:state}). $\mathcal{P}_u$ denotes the user-preferred attribute. $\mathcal{P}_{\mathrm{rej}}$ and $\mathcal{V}_{\mathrm{rej}}$ are the attributes and items rejected by the user, respectively.
The initial state $s_0$ is initialized by the user-specified attribute $p_0$, i.e., $s_0 = [[\{p_0\},\{\},\{\}],\mathcal{G}_u^{(0)}]$.

\subsubsection{\textbf{Action}}
According to the state $s_t$, the agent takes an action $a_t\in\mathcal{A}$, where $a_t$ can be selected from the candidate item set $\mathcal{V}_{\mathrm{cand}}^{(t)}$ to recommend items or from the candidate attribute set $\mathcal{P}_{\mathrm{cand}}^{(t)}$ to ask attributes. Following the path reasoning approach~\cite{kdd20-scpr}, we have
\begin{align}
    \mathcal{V}_{\mathrm{cand}}^{(t)} = \mathcal{V}_{\mathcal{P}_u^{(t)}}\setminus \mathcal{V}_{\mathrm{rej}}^{(t)},\quad
    \mathcal{P}_{\mathrm{cand}}^{(t)} = \mathcal{P}_{\scaleto{\mathcal{V}_{\mathrm{cand}}^{(t)}}{9pt}  }\setminus(\mathcal{P}_u^{(t)}\cup\mathcal{P}_{\mathrm{rej}}^{(t)}),
\end{align}
where $\mathcal{V}_{\mathcal{P}_u^{(t)}}$ is the set of item vertices directly connecting all $\mathcal{P}_u^{(t)}$ (i.e., items satisfying all the preferred attributes), and $\mathcal{P}_{\scaleto{\mathcal{V}_{\mathrm{cand}}^{(t)}}{9pt}}$ is the set of attribute vertices directly connecting to one of $\mathcal{V}_{\mathrm{cand}}^{(t)}$ (i.e., attributes belonging to at least one of the candidate items).

\subsubsection{\textbf{Transition}}
We consider that the current state $s_t$ will transition to the next state $s_{t+1}$ when the user responds to the action $a_t$. In specific, if CRS asks an attribute $p_t$ and the user accepts it, the next state $s_{t+1}$ will be updated by $\mathcal{P}_u^{(t+1)}=\mathcal{P}_u^{(t)}\cup p_t$. Conversely, if the user rejects the action $a_t$, $s_{t+1}$ will be updated by $\mathcal{P}_{\mathrm{rej}}^{(t+1)}=\mathcal{P}_{\mathrm{rej}}^{(t)}\cup a_t$ or $\mathcal{V}_{\mathrm{rej}}^{(t+1)}=\mathcal{V}_{\mathrm{rej}}^{(t)}\cup a_t$ for $a_t\in\mathcal{P}$ or $a_t\in\mathcal{V}$, respectively. As a result, the next state $s_{t+1}$ will be $[\mathcal{H}_u^{(t+1)},\mathcal{G}_u^{(t+1)}]$.

\subsubsection{\textbf{Reward}} 
Following previous MCR studies~\cite{wsdm20-ear,kdd20-scpr}, our environment contains five kinds of rewards, namely, (1) $r_{\mathrm{rec\_suc}}$, a strongly positive reward when the user accepts the recommended items,  (2) $r_{\mathrm{rec\_fail}}$, a negative reward when the user rejects the recommended items, (3) $r_{\mathrm{ask\_suc}}$, a slightly positive reward when the user accepts the asked attribute, (4) $r_{\mathrm{ask\_fail}}$, a negative reward when the user rejects the asked attribute, and (5) $r_{\mathrm{quit}}$, a strongly negative reward when reaching the maximum number of turns.

\subsection{Graph-enhanced State Representation}
\label{sec:state}

As we formulate conversational recommendation as a unified policy learning problem over a graph-based MDP environment, it is required to encode both the conversational and graph structural information into the latent distributed representations. In order to make use of the interrelationships among users, items, and attributes, we first adopt graph-based pre-training methods~\cite{transe,transh} to obtained node embeddings for all the nodes in the full graph $\mathcal{G}$.

\subsubsection{\textbf{Dynamic Weighted Graph Construction}}
As shown in Figure~\ref{method}, we represent the current state of the graph-based MDP environment as a dynamic weighted graph. 
Formally, we denote an undirected weighted graph as $\mathcal{G}=(\mathcal{N},\bm{A})$, with the node $n_i\in\mathcal{N}$, the adjacency matrix element $\bm{A}_{i,j}$ denoting the weighted edges between nodes $n_i$ and $n_j$. In our case, given the user $u$, we denote the dynamic graph at timestep $t$ as $\mathcal{G}_u^{(t)}=(\mathcal{N}^{(t)},\bm{A}^{(t)})$:
\begin{equation}
    \mathcal{N}^{(t)} = \{u\}\cup\mathcal{P}_u^{(t)}\cup\mathcal{P}_{\mathrm{cand}}^{(t)}\cup\mathcal{V}_{\mathrm{cand}}^{(t)}
\end{equation}
\begin{equation}
    \bm{A}^{(t)}_{i,j} =
  \begin{cases}
    w_v^{(t)},     & \quad \text{if } n_i = u, n_j \in \mathcal{V} \\
    1,       & \quad \text{if } n_i \in \mathcal{V}, n_j \in \mathcal{P} \\
    0,       & \quad \text{otherwise}
  \end{cases} 
\label{eq:a}
\end{equation}
where $w_v^{(t)}$ is a scalar indicating the recommendation score of the item $v$ in the current state. In order to incorporate the user preference as well as the correlation between the asked attributes and the items, such weight $w_v^{(t)}$ is calculated as
\begin{equation}
    w_v^{(t)} = \sigma\Bigl(e_u^\top e_v + \sum\nolimits_{\scaleto{p\in\mathcal{P}_u^{(t)}}{8pt}} e_v^\top e_p - \sum\nolimits_{\scaleto{p\in\mathcal{P}_{\mathrm{rej}}^{(t)}\cap\mathcal{P}_v^{(t)}}{8pt}} e_v^\top e_p\Bigr),
\label{item_weight}
\end{equation}
where $\sigma(\cdot)$ denotes the sigmoid function, $e_u$, $e_v$, and $e_p$ are the embeddings of the user, item, and attribute, respectively. 

\subsubsection{\textbf{Graph-based Representation Learning}}
In order to comprehensively take advantage of the correlation information among the involved user, items, and attributes from the connectivity of the graph, we employ a graph convolutional network (GCN)~\cite{iclr17-gcn} to refine the node representations with structural and relational knowledge. The representations of the node $n_i$ in the $(l+1)$-th layer can be computed by:
\begin{equation}
    e_i^{(l+1)} = \text{ReLU}\left(\sum\nolimits_{j\in \mathcal{N}_i}\bm{\Lambda}_{i,j}\bm{W}_{l}e_j^{(l)} + \bm{B}_le_i^{(l)}\right),
\end{equation}
where $\mathcal{N}_i$ denotes the neighboring indices of the node $n_i$, $\bm{W}_{l}$ and $\bm{B}_{l}$ are trainable parameters representing the transformation from neighboring nodes and the node $n_i$ itself, and $\bm{\Lambda}$ is a normalization adjacent matrix as $\bm{\Lambda} = \bm{D}^{-\frac{1}{2}}\bm{A}\bm{D}^{-\frac{1}{2}}$ with $\bm{D}_{ii} =\sum_j \bm{A}_{i,j} $.

\subsubsection{\textbf{Sequential Representation Learning}}
Apart from the interrelationships among the involved user, items, and attributes, the CRS is also expected to model the conversation history in the current state. Unlike previous studies~\cite{wsdm20-ear,kdd20-scpr} that adopt heuristic features for conversation history modeling, we employ Transformer encoder~\cite{transformer} for capturing the sequential information of the conversation history as well as attending the important information for deciding the next action. As described in \cite{transformer}, each Transformer layer consists of three components: (i) The layer normalization is defined as LayerNorm$(\cdot)$. (ii) The multi-head attention is defined as MultiHead$(\bm{Q}, \bm{K}, \bm{V})$, where $\bm{Q}, \bm{K}, \bm{V}$ are query, key, and value, respectively. (iii) The feed-forward network with ReLU activation is defined as FFN$(\cdot)$. Take the $l$-th layer for example: 
\begin{align}
    \bm{X}^* &= \text{MultiHead}(\bm{X}^{(l)},\bm{X}^{(l)},\bm{X}^{(l)}), \\
    \bm{X}^{(l+1)} &=\text{LayerNorm}(\text{FFN}(\bm{X}^*) + \bm{X}^{(l)}),
\end{align}
where $\bm{X}\in\mathbb{R}^{L\times d}$ denotes the embeddings, and $L$ is the sequence length. In our case, the input sequence $\bm{X}^{(0)}$ is the accepted attributes $\mathcal{P}_u^{(t)}$ in the current conversation history with the learned graph-based representation $\{e_p^{(L_g)}:p\in\mathcal{P}_u^{(t)}\}$, where $L_g$ is the number of layers in GCN. After the sequential learning with $L_s$ Transformer layers, we can aggregate the information learned from both the graph and the conversation history by a mean pooling layer to obtain the state representation of $s_t$:
\begin{equation}
    f_{\theta_S}(s_t) = \text{MeanPool}(\bm{X}^{L_s}).
\label{state_rep}
\end{equation}
For simplicity, we denote the learned state representation of $s_t$ as $f_{\theta_S}(s_t)$, where $\theta_S$ is the set of all network parameters for state representation learning, including GCN and Transformer layers.

\subsection{Action Selection Strategy}
\label{sec:action}
A large action search space will harm the performance of the policy learning to a great extent~\cite{kdd20-scpr}. Thus, it attaches great importance to handle the overwhelmingly large action space in UCRPL.
To this end, we propose two simple strategies to improve the sample efficiency for candidate action selection. 

\noindent \textbf{Preference-based Item Selection}. In general, for candidate items to be recommended, we can consider only the action of making recommendations from a small number of candidate items that fit the user preference the most, since users are not likely to be interested in all items. To achieve this, we select top-$K_v$ candidate items from $\mathcal{V}_{\mathrm{cand}}^{(t)}$ into the candidate action space $\mathcal{A}_t$ at each timestep $t$, which is ranked by the recommendation score $w_v^{(t)}$ in Eq.(\ref{item_weight}).  

\noindent \textbf{Weighted Entropy-based Attribute Selection}. Whereas for candidate attributes to be asked, the expected one is supposed to be able to not only better eliminate the uncertainty of candidate items, but also encode the user preference. Inspired by~\cite{kdd20-scpr}, we adopt weighted entropy as the criteria to prune candidate attributes:
\begin{align}
    w_p^{(t)} &= - \text{prob}(p^{(t)})\cdot\log(\text{prob}(p^{(t)})),\\
    \text{prob}(p^{(t)}) &= \sum\nolimits_{v\in\mathcal{V}_{\mathrm{cand}}^{(t)}\cap\mathcal{V}_p^{(t)}}w_v^{(t)} \Big/ \sum\nolimits_{v\in\mathcal{V}_{\mathrm{cand}}^{(t)}}w_v^{(t)},
\end{align}
where $\mathcal{V}_p$ denotes the items that have the attribute $p$. Similar to item selection, we also select top-$K_p$ candidate attributes from $\mathcal{P}_{\mathrm{cand}}^{(t)}$ into $\mathcal{A}_t$ based on the weighted entropy score $w_p^{(t)}$.

\begin{algorithm}[t]
\small
\caption{Unified Conversational Recommender}
\label{algo}
\KwIn{$\{\mathbf{e}_i\}_{i\in \mathcal{N}}$; $\mathcal{D}$;  $\tau$; $\epsilon$; $\gamma$; $K$; $T$; $K_v$; $K_p$;} 
\KwOut{ $\theta_S$;  $\theta_Q$; }
Initialize all parameters: $\theta_S$, $\theta_Q$, $\theta_Q' \leftarrow \theta_Q$\;

\For{$\mathit{episode} = 1, 2, \ldots , N$}{

User $u$ starts the conversation by specifying an attribute $p_0$\;

Update: $\mathcal{P}_u^{(0)} = \{p_0\}, \quad$ $\mathcal{P}_{\mathrm{rej}}^{(0)}=\{\}, \quad$ $\mathcal{V}_{\mathrm{rej}}^{(0)}=\{\}$, $\mathcal{H}_u^{(0)} = [\mathcal{P}_u^{(0)},\mathcal{P}_{\mathrm{rej}}^{(0)},\mathcal{V}_{\mathrm{rej}}^{(0)}], \quad$ $s_0 = [\mathcal{H}_u^{(0)}, \mathcal{G}_u^{(0)}]$\;
Get candidate action space $\mathcal{A}_0$ via Action Selection\;
		
\For{turn $t = 0, 1, \ldots , T-1$}{
	
	Get state representation $f_{\theta_S}(s_t)$ via Eq.(\ref{state_rep})\;
	Select an action $a_t$ by $\epsilon$-greedy w.r.t Eq.(\ref{q-value})\;
	
	Receive reward $r_t$\;
	Update the next state $s_{t+1} = \mathcal{T}(s_t,a_t)$\;
	Get $\mathcal{A}_{t+1}$ via Action Selection\;
	Store $(s_t, a_t, r_t, s_{t+1}, \mathcal{A}_{t+1})$ to buffer $\mathcal{D}$\;
	Sample mini-batch of $(s_t, a_t, r_t, s_{t+1}, \mathcal{A}_{t+1})$ w.r.t Eq.(\ref{per})\;
	Compute the target value $y_t$ via Eq.~(\ref{target_value})\;
	Update $\theta_S$, $\theta_Q$ via SGD w.r.t the loss function Eq.(\ref{loss})\;
	Update $\theta_Q'$ via Eq.(\ref{double}) \;
}
}
\end{algorithm}

\subsection{Deep Q-Learning Network}
After obtaining the graph-enhanced state representation and the candidate action space, we introduce the deep Q-learning network (DQN)~\cite{nature-dqn} to conduct the unified conversational recommendation policy learning. We further implement some techniques to enhance and stabilize the training of DQN. The training procedure of the Unified Conversational Recommender is presented in Algorithm~\ref{algo}.

\subsubsection{\textbf{Dueling Q-Network}}
Following the standard assumption that delayed rewards are discounted by a factor of $\gamma$ per timestep, we define
the Q-value $Q(s_t, a_t)$ as the expected reward based on the state $s_t$ and the action $a_t$. 
As shown in the rightmost part of Figure~\ref{method}, the dueling Q-network~\cite{icml16-dueling} employs two deep neural networks to compute the value function $f_{\theta_V}(\cdot)$ and advantage function $f_{\theta_A}(\cdot)$, respectively. Then the Q-function can be calculated by:
\begin{equation}\label{q-value}
    Q(s_t, a_t) = f_{\theta_V}(a_t) + f_{\theta_A}(f_{\theta_S}(s_t), a_t),
\end{equation}
where $f_{\theta_V}(\cdot)$ and $f_{\theta_A}(\cdot)$ are two separate multi-layer perceptions with parameters $\theta_V$ and $\theta_A$, respectively, and let $\theta_Q = \{\theta_V, \theta_A\}$.

The optimal Q-function $Q^*(s_t, a_t)$, which has the maximum expected reward achievable by the optimal policy $\pi^*$, follows the Bellman equation~\cite{bellman} as:
\begin{equation}
    Q^*(s_t, a_t) = \mathbb{E}_{s_{t+1}}\Bigl[r_t+\gamma \max_{a_{t+1}\in\mathcal{A}_{t+1}}\! Q^*(s_{t+1},a_{t+1})|s_t,a_t\Bigr].
\end{equation}

\subsubsection{\textbf{Double Q-Learning with Prioritized Experience Replay}}
During each episode in the MCR process, at each timestep $t$, the CRS agent obtains the current state representation $f_{\theta_S}(s_t)$ via the graph-enhanced state representational learning described in Section~\ref{sec:state}. Then the agent selects an action $a_t$ from the candidate action space $\mathcal{A}_t$, which is obtained via the action selection strategies described in Section~\ref{sec:action}. Here we incorporate $\epsilon$-greedy method to balance the exploration and exploitation in action sampling (i.e., select a greedy action based on the max Q-value with probability $1-\epsilon$, and a random action with probability $\epsilon$).

Then, the agent will receive the reward $r_t$ from the user's feedback. According to the feedback, the current state $s_t$ transitions to the next state $s_{t+1}$, and the candidate action space $\mathcal{A}_{t+1}$ is updated accordingly. The experience $(s_t, a_t, r_t, s_{t+1}, \mathcal{A}_{t+1})$ is then stored into the replay buffer $\mathcal{D}$. To train DQN, we sample mini-batch of experiences from $\mathcal{D}$, and minimize the following loss function:
\begin{align}
    \mathcal{L}(\theta_Q, \theta_S) &= \mathbb{E}_{(s_t,a_t,r_t,s_{t+1},\mathcal{A}_{t+1}){\sim}\mathcal{D}}\bigl[(y_t {-} Q(s_t,a_t;\theta_Q,\theta_S))^2\bigr], \label{loss}\\
    y_t &= r_t + \gamma \max_{a_{t+1}\in\mathcal{A}_{t+1}} Q(s_{t+1},a_{t+1};\theta_Q,\theta_S), \label{eq:yt}
\end{align}
where $y_t$ is the target value based on the currently optimal $Q^*$. 

To alleviate the overestimation bias problem in conventional DQN, we adopt Double Q-learning~\cite{aaai16-doubledqn}, which employs a target network $Q'$ as a periodic copy from the online network. The target value of the online network is then changed to:
\begin{equation}
    y_t = r_t + \gamma Q'\bigl(s_{t+1}, \argmax_{a_{t+1}\in\mathcal{A}_{t+1}} Q(s_{t+1},a_{t+1};\theta_Q,\theta_S);\theta_{Q'},\theta_S\bigr),
\label{target_value}
\end{equation}
where $\theta_{Q'}$ denotes the parameter of the target network, which is updated by the soft assignment as:
\begin{equation}
    \theta_{Q'} = \tau\theta_{Q} + (1-\tau)\theta_{Q'},
\label{double}
\end{equation}
where $\tau$ is the update frequency.

In addition, the conventional DQN samples uniformly from the replay buffer. In order to sample more frequently those important transitions from which there is much to learn, we employ prioritized replay~\cite{iclr16-per} as a proxy for learning potential, which samples transitions with probability $\delta$ relative to the absolute TD error:
\begin{equation}
    \delta \propto \Bigl|r_{t+1} + \gamma Q'\bigl(s_{t+1},\argmax_{a_{t+1}\in\mathcal{A}_{t+1}} Q(s_{t+1},a_{t+1})\bigr) - Q(s_t,a_t)\Bigr|.
\label{per}
\end{equation}

\subsubsection{\textbf{Model Inference}}
With the learned UNICORN model, given a user and his/her conversation history, we follow the same process to obtain the candidate action space and the current state representation, and then decide the next action according to max Q-value in Eq.(\ref{q-value}). 
If the selected action points to an attribute, the system will ask the user's preference on the attribute. Otherwise, the system will recommend top-$K$ items with the highest Q-value to the user.

\section{Experiment}

\subsection{Dataset}
We evaluate the proposed method, UNICORN, on four existing multi-round conversational recommendation benchmark datasets, and further conduct evaluation on a real-world E-Commerce platform. The statistics of these datasets are presented in Table~\ref{dataset}.

\begin{itemize}[leftmargin=*,topsep=4pt]
    \item \textbf{LastFM} and \textbf{Yelp}. The LastFM dataset is used for evaluation on music artist recommendation, while the Yelp dataset is for business recommendation. \citet{wsdm20-ear} manually categorize the original attributes in LastFM into 33 coarse-grained groups, and build a 2-layer taxonomy with 29 first-layer categories for Yelp. 
    \item \textbf{LastFM*} and \textbf{Yelp*}. \citet{kdd20-scpr} consider that it is not realistic to manually merge attributes for practical applications, so they adopt original attributes to reconstruct these two datasets. 
    For a fair comparison, we adopt both versions for experiments.
    \item \textbf{E-Commerce}. In order to evaluate the proposed method in real-world E-Commerce scenario, we collect the dataset by sampling logs from Taobao, the largest E-Commerce platform in China, in the same way as LastFM and Yelp. We randomly sample 30,000 items from the Women\_Dress category and a certain number of users who have interacted with these items. And we treat the concatenation of the product property and the property value as the attribute to be asked, e.g., ``Color=Red'', where ``Color'' is a product property and ``Red'' is a property value. Following the MCR data construction described in \citet{wsdm20-ear}, we obtain the E-Commerce dataset with the statistics shown in Table~\ref{dataset}. As for  graph construction, we consider two kinds of relations, including ``purchased\_by (item$\rightarrow$ user)'' and ``belong\_to (attribute$\rightarrow$ item)''. Compared with the imbalanced ratio of \#items:\#attributes at LastFM (0.88:1) and Yelp (119:1), the E-Commerce dataset preserves a large number of items as well as attributes, which is more in line with real-world applications in E-Commerce. 
\end{itemize}

\begin{table}
\setlength{\abovecaptionskip}{2pt}   
\setlength{\belowcaptionskip}{2pt}
\centering
  \caption{Summary statistics of datasets.}
\begin{adjustbox}{max width=\linewidth}
\setlength{\tabcolsep}{1mm}{
\begin{tabular}{lrrrrr}
\toprule 
& LastFM & LastFM* & Yelp & Yelp* & E-Com. \\
\midrule 
 \#Users & 1,801 & 1,801 & 27,675 & 27,675 & 26,430 \\
 \#Items & 7,432 & 7,432 & 70,311 & 70,311 & 29,428 \\
 \#Interactions & 76,693 & 76,693 & 1,368,606 & 1,368,606 & 748,533 \\
 \#Attributes & 33 & 8,438 & 29 & 590 & 1,413 \\
\midrule 
\#Entities & 9,266 & 17,671 & 98,605 & 98,576 & 57,271 \\
 \#Relations & 4 & 4 & 3 & 3 & 2 \\
 \#Triplets & 138,215 & 228,217 & 2,884,567 & 2,533,827 & 2,024,962 \\
\bottomrule
\label{dataset}
\end{tabular}}
\end{adjustbox}
\vspace{-0.5cm}
\end{table}

\begin{table*}
\centering
  \caption{Experimental Results. $^\dagger$ indicates statistically significant improvement ($p<0.01$) over all baselines. hDCG stands for hDCG@(15,10). SR and hDCG are the higher the better, while AT is the lower the better.}
\begin{adjustbox}{max width=\textwidth}
\setlength{\tabcolsep}{1mm}{
\begin{tabular}{lccccccccccccccc}
\toprule 
& \multicolumn{3}{c} { LastFM } & \multicolumn{3}{c} { LastFM*} & \multicolumn{3}{c} { Yelp} & \multicolumn{3}{c} { Yelp* } & \multicolumn{3}{c} { E-Commerce}\\
\cmidrule(lr){2-4} \cmidrule(lr){5-7} \cmidrule(lr){8-10} \cmidrule(lr){11-13} \cmidrule(lr){14-16}
& SR@15 & AT &hDCG & SR@15 & AT&hDCG & SR@15 & AT&hDCG & SR@15 & AT&hDCG& SR@15 & AT &hDCG\\
\midrule 
Abs Greedy & 0.222 & 13.48 &0.073& 0.635 & 8.66 &0.267& 0.264 & 12.57 &0.145& 0.189 & 13.43 &0.089& 0.273&12.19&0.138 \\
Max Entropy & 0.283 & 13.91 &0.083& 0.669 & 9.33 &0.269& 0.921 & 6.59 &0.338& 0.398 & 13.42&0.121&0.328&12.98& 0.112\\
CRM~\cite{sigir18-crm} & 0.325 & 13.75 &0.092& 0.580 & 10.79 &0.224& 0.923 & 6.25 &0.353& 0.177 & 13.69&0.070&0.294&12.11&0.146 \\
EAR~\cite{wsdm20-ear} & 0.429 & 12.88 &0.136& 0.595 & 10.51 &0.230& 0.967 & 5.74 &0.378& 0.182 & 13.63 &0.079&0.381&\underline{11.48}&0.161\\
SCPR~\cite{kdd20-scpr} & $\underline{0.465}$ & $\underline{12.86}$ & \underline{0.139} &$\underline{0.709}$ & $\underline{8.43}$ & \underline{0.317} &$\underline{0.973}$ & $\underline{5.67}$ & \underline{0.382} &$\underline{0.489}$ & $\underline{12.62}$ &\underline{0.159} &\underline{0.518} &12.32&\underline{0.168}\\
\midrule 
\textbf{UNICORN} & $\mathbf{0.535}$$^\dagger$ & $\mathbf{11.82}$$^\dagger$ &\textbf{0.175}$^\dagger$& $\mathbf{0 . 788}$$^\dagger$ & $\mathbf{7 . 58}$$^\dagger$ &\textbf{0.349}$^\dagger$& $\mathbf{0 . 9 8 5}$$^\dagger$ & $\mathbf{5 . 33}$$^\dagger$ &\textbf{0.397}$^\dagger$& $\mathbf{0 . 520}$$^\dagger$ & $\mathbf{11 .31}$$^\dagger$&\textbf{0.203}$^\dagger$&\textbf{0.602$^\dagger$}&\textbf{10.45}$^\dagger$ &\textbf{0.217}$^\dagger$\\
\bottomrule
\label{result}
\end{tabular}}
\end{adjustbox}
\vspace{-0.3cm}
\end{table*}

\subsection{Experimental Settings}
\subsubsection{\textbf{User Simulator}}\label{sec:user_sim} 
Due to the interactive nature of MCR, it needs to be trained and evaluated by interacting with users. Following the user simulator adopted in~\cite{sigir18-crm,wsdm20-ear}, we simulate a conversation session for each observed user-item interaction pair $(u,v)$. We regard the item $v$ as the ground-truth target item and treat its attribute set $\mathcal{P}_v$ as the oracle set of attributes preferred by the user $u$ in this conversation session. The session is initialized by the simulated user who specifies a certain attribute randomly chosen from $\mathcal{P}_v$. Then the session follows the process of ``System Ask, User Respond''~\cite{cikm18-saur} as described in Section~\ref{sec:method}. 

\subsubsection{\textbf{Baselines}}
We compare the proposed method with several state-of-the-art methods on MCR as follows:
\begin{itemize}[leftmargin=*,topsep=4pt]
    \item \textbf{Max Entropy}. This is a rule-based strategy to decide the next action, where the CRS selects an attribute to ask based on the maximum entropy within the current state, or chooses to recommend the top-ranked items with a certain probability~\cite{wsdm20-ear}.
    \item \textbf{Abs Greedy}~\cite{kdd16-convbandit}. This method only performs recommendation actions and updates itself until the CRS makes a successful recommendation or exceeds the maximum turns of conversation.
    \item \textbf{CRM}~\cite{sigir18-crm}. This method is originally designed for single-round conversational recommendation, which employs policy gradient~\cite{nips99-pg} to learn the policy deciding when and which attributes to ask. We follow~\cite{wsdm20-ear} to adapt this method into MCR scenario.
    \item \textbf{EAR}~\cite{wsdm20-ear}. A three-stage method is proposed to enhance the interaction between the conversation and recommendation components with a similar RL framework as CRM.
    \item \textbf{SCPR}~\cite{kdd20-scpr}. This is the state-of-the-art method on MCR setting, which leverages interactive path reasoning on the graph to prune off candidate attributes and adopts the DQN~\cite{nature-dqn} framework to determine when to ask or recommend.
\end{itemize}

\subsubsection{\textbf{Evaluation Metrics}}
Following previous studies on multi-round conversational recommendation~\cite{wsdm20-ear,kdd20-scpr}, we adopt success rate at the turn $t$ (SR@$t$)~\cite{sigir18-crm} to measure the cumulative ratio of successful conversational recommendation by the turn $t$. Besides, average turn (AT) is adopted to evaluate the average number of turns for all sessions (if the conversation session reaches the maximum turn $T$, the turn for such session is also counted as $T$). The higher SR@$t$ indicates a better performance of the CRS at a turn $t$, while the lower AT means an overall higher efficiency. 

Although SR@$t$ and AT are widely adopted for the evaluation of CRS, both of them are not sensitive to the rank order in the recommendation results and only consider a certain aspect of the performance of CRS. 
In order to conduct a comprehensive evaluation of CRS, we propose to extend the normalized discounted cumulative gain (NDCG@$K$) to be a two-level hierarchical version, namely \textbf{hNDCG@($\bm{T,K}$)}, with the definition as follows:
\begin{equation}
    \begin{aligned}
    hDCG@(T,K) =& \sum\nolimits_t^T\sum\nolimits_k^Kr(t,k)\left[\frac{1}{\log_2(t+2)}\right.\\
    &\left. +\left(\frac{1}{\log_2(t{+}1)}-\frac{1}{\log_2(t{+}2)}\right)\frac{1}{\log_2(k{+}1)} \right],
\end{aligned}
\end{equation}
where $T$ and $K$ represent the number of conversation turns and recommended items in each turn, $r(t,k)$ denotes the relevance of the result at the turn $t$ and position $k$.
Then hNDCG@($T,K$) can be drawn from hDCG@($T,K$) with the same way as the original NDCG@$K$. Since there is only one target item for each session in MCR, we simply use hDCG@($T,K$) for evaluation. The intuition behind hNDCG@($T,K$) is that the less number of turns of a successful session is favorable for the CRS, while the target item is expected to be ranked higher in the recommendation list at the success turn.

\begin{figure}
\setlength{\abovecaptionskip}{2pt}   
\setlength{\belowcaptionskip}{2pt}
\centering
\includegraphics[width=0.48\textwidth]{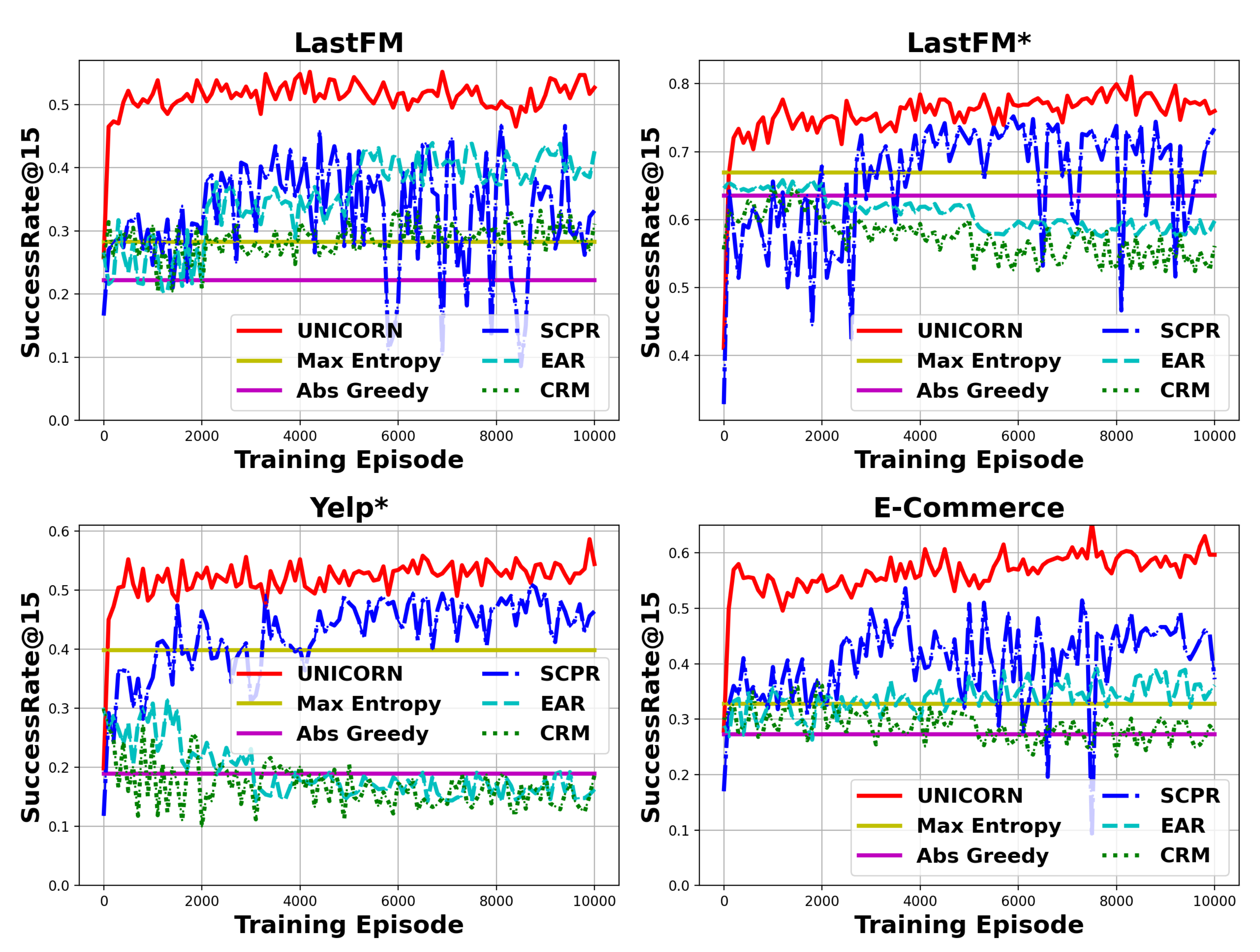}
\caption{Test Performance at Different Training Episodes.}
\label{converge}
\vspace{-0.3cm}
\end{figure}

\subsubsection{\textbf{Implementation Details}}
Following~\cite{kdd20-scpr}, we split the E-Commerce dataset by 7:1.5:1.5 for training, validation, and testing, and set the size $K$ of the recommendation list as 10, the maximum turn $T$ as 15. 
We adopt TransE~\cite{transe} from OpenKE~\cite{openke} to pretrain the node embeddings in the constructed graph with the training set. 
We adopt the user simulator described in Section~\ref{sec:user_sim} to interact with the CRS for online training the model using the validation set.  For all implemented methods, we conduct the online training for 10,000 episodes. To maintain a fair comparison with other baselines~\cite{wsdm20-ear,kdd20-scpr}, we adopt the same reward settings to train the proposed model: $r_{\mathrm{rec\_suc}}$=1, $r_{\mathrm{rec\_fail}}$=-0.1, $r_{\mathrm{ask\_suc}}$=0.01, $r_{\mathrm{ask\_fail}}$=-0.1, $r_{\mathrm{quit}}$=-0.3. The hyper-parameters are empirically set as follows: The embedding size and the hidden size are set to be 64 and 100. The numbers of GCN layers $L_g$ and Transformer layers $L_s$ are set to be 2 and 1, respectively. The numbers of selected candidate attributes $K_p$ and items $K_v$ are set to be 10. 
During the training procedure of DQN, the size of experience replay buffer is 50,000, and the size of mini-batch is 128. The learning rate and the $L_2$ norm regularization are set to be 1e-4 and 1e-6, with Adam optimizer. The discount factor $\gamma$ and the update frequency $\tau$ are set to be 0.999 and 0.01.

\subsection{Performance Comparison}
\subsubsection{\textbf{Overall Performance}}

\begin{table*}
\setlength{\abovecaptionskip}{2pt}   
\setlength{\belowcaptionskip}{2pt}
\centering
  \caption{Ablation Study. SR and hDCG are the higher the better, while AT is the lower the better.}
\begin{adjustbox}{max width=\textwidth}
\setlength{\tabcolsep}{1mm}{
\begin{tabular}{lccccccccccccccc}
\toprule 
& \multicolumn{3}{c} { LastFM } & \multicolumn{3}{c} { LastFM*} & \multicolumn{3}{c} { Yelp} & \multicolumn{3}{c} { Yelp* } & \multicolumn{3}{c} { E-Commerce}\\
\cmidrule(lr){2-4} \cmidrule(lr){5-7} \cmidrule(lr){8-10} \cmidrule(lr){11-13} \cmidrule(lr){14-16}
& SR@15 & AT &hDCG & SR@15 & AT &hDCG & SR@15 & AT &hDCG & SR@15 & AT &hDCG& SR@15 & AT &hDCG\\
\midrule 
\textbf{UNICORN} & \textbf{0.535} & \textbf{11.82} &\textbf{0.175}& \textbf{0.788} & \textbf{7.58} &\textbf{0.349}& \textbf{0.985} & \textbf{5.33} &\textbf{0.397}& \textbf{0.520} & \textbf{11.31}&\textbf{0.203}&\textbf{0.602}&\textbf{10.45} &\textbf{0.217}\\
\midrule
(a) - w/o Transformer&0.478&12.17&0.147&0.733&7.78&0.320&0.958&6.18&0.370&0.470&11.62&0.167&0.545&11.23&0.178\\
(b) - w/o GCN&0.440&12.74&0.139&0.744&7.62&0.327&0.978&5.48&0.394&0.481&11.45&0.171&0.554&11.07&0.183\\
\midrule
(c) - Random Embeddings&0.310&13.31&0.096&0.665&9.24&0.291&0.905&7.66&0.318&0.196&13.95&0.049&0.220&13.55&0.069\\
(d) - FM Embeddings&0.465 & 12.26&0.144 & 0.748 & 7.68&0.328 & 0.982 &5.50& 0.394& 0.496 & 12.25&0.161&0.579&10.89&0.202\\
\midrule 
(e) - Preference-based Attr. Sel.&0.467&12.33&0.129&0.712&8.24&0.303&0.978&5.74&0.385&0.489&11.60&0.171&0.569&11.06&0.206\\
(f) - Entropy-based Attr. Sel.&0.507&11.96&0.165&0.756&7.63&0.324&0.978&5.71&0.388&0.502&11.33&0.192&0.585&10.69&0.214\\
(g) - w/o Attribute Selection&0.487&12.12&0.151&0.604&9.88&0.258&0.942&6.80&0.357&0.220&12.97&0.067&0.434&12.01&0.127\\
(h) - w/o Item Selection&0.150&13.83&0.055&0.638&8.85&0.294&0.780&9.16&0.276&0.158&13.60&0.054&0.144&13.85&0.056\\
\bottomrule
\end{tabular}}
\end{adjustbox}
\label{ablation}
\end{table*}

Table~\ref{result} shows the performance comparison between the proposed method, UNICORN, and all baselines across five datasets. In general, UNICORN outperforms all the baselines by achieving a significantly higher success rate and less average turn, which is also comprehensively validated by the improvements on hDCG. As for the real-world E-Commerce dataset, SCPR outperforms EAR and CRM, whose performance is largely affected by the large action space in E-Commerce dataset. Despite the effectiveness of SCPR to handle the issue of large action space in CRS, its performance in such a real-world application is still restricted by the separation of its conversation and recommendation components. Our UNICORN not only enables the conversation and recommendation to be mutually enhanced during the training process, but also attains an effective sample efficiency with the proposed action selection strategies. This leads to a substantial margin from these baselines, about 18\% for SR@15, 2 turns for AT, and 30\% for hDCG. Detailed analyses can be found as follows.

\subsubsection{\textbf{Training Efficiency}}

Figure~\ref{converge} shows the test performance curves of different methods. Since Max Entropy and Abs Greedy are unsupervised methods, their curves are presented as two lines for comparisons. It can be clearly observed that UNICORN is trained more stably and requires fewer training episodes (i.e., interaction data) to achieve a better performance than other strong baselines. Among these baselines, the curve of SCPR is the most vibrant, since it only considers the policy of when to ask or recommend, while the decisions of question-asking and recommendation are made by two separated components. As for EAR and CRM, due to the large action space in the last three datasets, there is no much performance increase during the online training process, even getting worse. These results demonstrate the efficiency and effectiveness of the proposed unified policy learning for CRS. 

\subsubsection{\textbf{Comparison at Different Conversation Turns}}

\begin{figure}
\setlength{\abovecaptionskip}{2pt}   
\setlength{\belowcaptionskip}{2pt}
\centering
\includegraphics[width=0.48\textwidth]{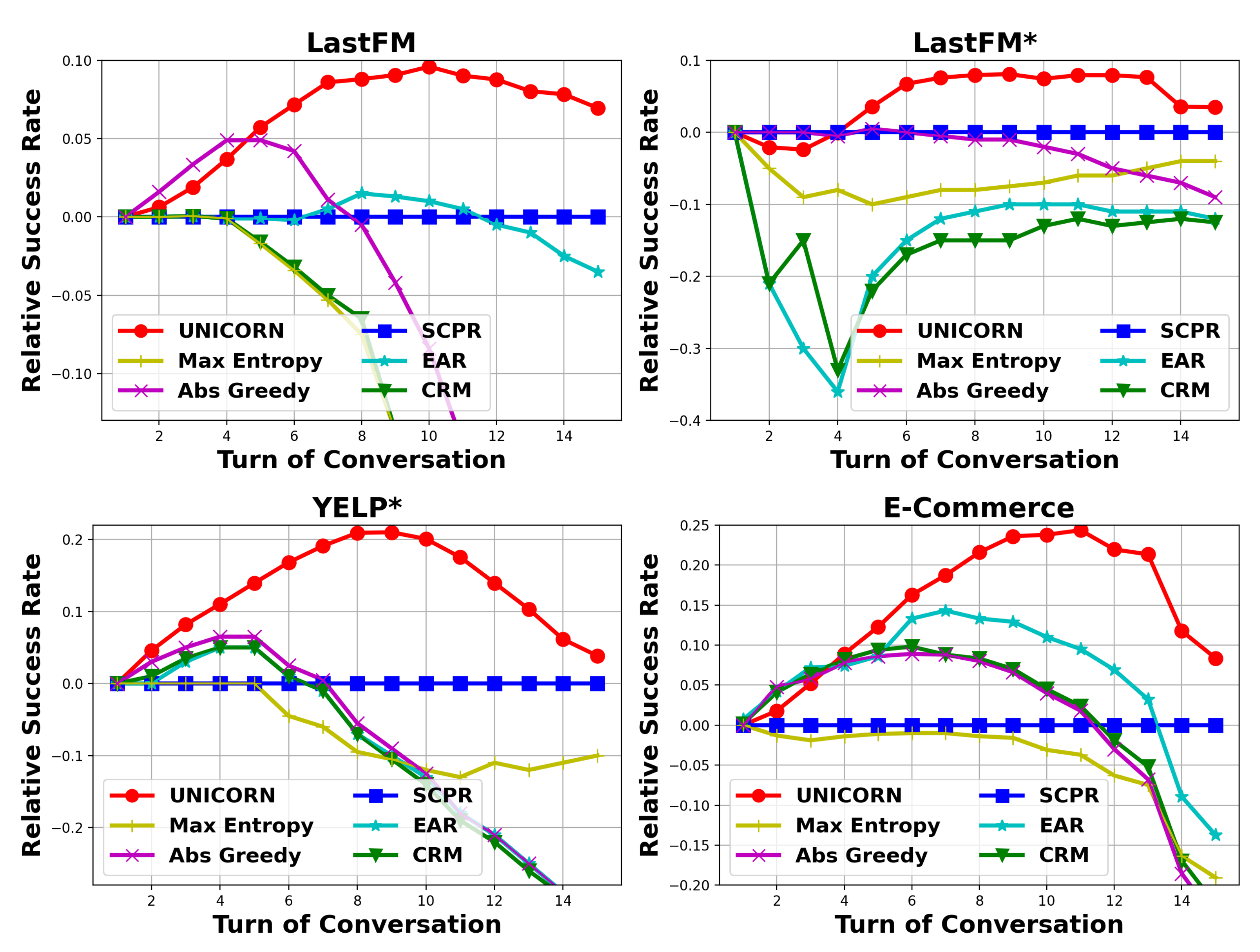}
\caption{Comparisons at Different Conversation Turns.}
\label{diff_turn}
\vspace{-0.3cm}
\end{figure}

Besides SR@15, we also present the performance comparison of success rate at each turn (SR@$t$) in Figure~\ref{diff_turn}. In order to better observe the differences among different methods, we report the relative success rate compared with the state-of-the-art baseline SCPR. For example, the line of $y=0$ represents the curve of Success Rate* for SCPR against itself. 
There are several notable observations as follows: 

(i) The proposed UNICORN substantially and consistently outperforms these baselines across all the datasets and almost each turn in the conversation session. 

(ii) Due to the nature of greedy recommendation approach (Abs Greedy), it may successfully hit target items at the early stage of the conversation, leading to a relatively strong performance at the first few turns, but the performance falls quickly as the turn increases. 

(iii) UNICORN achieves an outstanding performance in the middle stage of the conversation, where there are still a large number of candidate items and attributes to be pruned. This shows the strong scalability of UNICORN to effectively handle large candidate action space in different situations. 

(iv) The performance of SCPR is getting closer to UNICORN at the latter stage of the conversation, as the candidate item and attribute set is getting smaller and the task becomes easier. 

(v) EAR and CRM share similar performance as Abs Greedy in those datasets with a large candidate attribute set, i.e., Yelp* and E-Commerce, indicating their policy learning is merely working when encountering a large action space.

\begin{figure*}
\setlength{\abovecaptionskip}{2pt}   
\setlength{\belowcaptionskip}{2pt}
\centering
\includegraphics[width=0.95\textwidth]{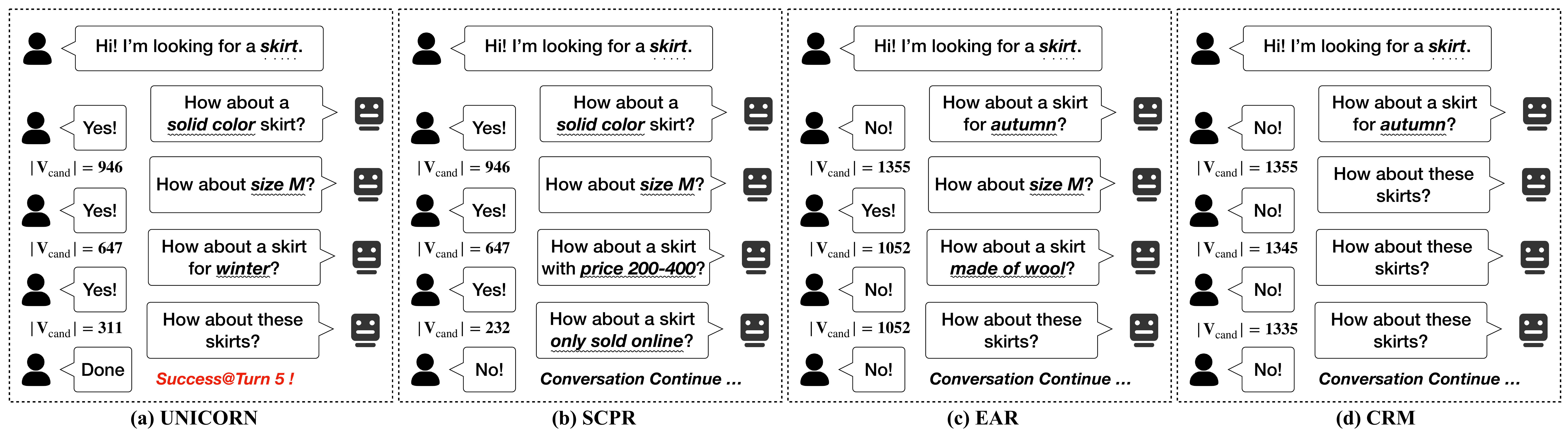}
\caption{Sample conversations generated by UNICORN, SCPR, EAR, and CRM. Due to the space limit, we only show the conversation at the first five turns. $|V_\mathrm{cand}|$ denotes the number of candidate items at the current turn.}
\label{case}
\vspace{-0.1cm}
\end{figure*}

\subsection{Ablation Study}

\subsubsection{\textbf{Model Components}}
We first evaluate the model components for the state representation learning, including the graph-based representation learning and the sequential representation learning. 
The ablation studies are presented in the first part in Table~\ref{ablation} (row (a-b)). 
We observe that the performance on 4 out of 5 datasets suffers a larger decrease by discarding sequential learning (row (a)) than discarding graph-based learning (row (b)). We attribute this to two reasons: (i) The pretrained TransE embeddings have already encoded certain graph-based knowledge into the node representations. (ii) Since there are relatively larger number of attributes in these four datasets than LastFM, it is more important to model the interrelationships among diverse user-accepted attributes for capturing user preferences.

\subsubsection{\textbf{Pre-training Embeddings}}
The second part in Table~\ref{ablation} (row (c-d)) presents the results that we replace pretrained TransE embeddings with randomly initialized embeddings (row (c)) and pretrained embeddings from FM model described in \citet{wsdm20-ear} (row (d)). We observe a substantial performance gap between UNICORN with randomly initialized embeddings and that with pretrained embeddings, either TransE or FM. This result shows that pretraining is necessary and simply online training with  interactive conversations is far from efficient and sufficient for training a CRS with limited interaction data. 
In addition, using TransE embeddings further improves the performance from FM embeddings to a great extent, which also validates the advantages of graph-enhanced RL.

\subsubsection{\textbf{Action Selection Strategies}}
The last part in Table~\ref{ablation} (row (e-h)) presents the results that we replace or discard the proposed action selection strategies. One alternative attribute selection strategy is to adopt the same strategy as the preference-based item selection by changing the object from items to attributes. Another one is to use the original maximum entropy function~\cite{wsdm20-ear}. The results (row (e,f)) show that the performance of UCRPL suffers a noticeable decrease when adopting the preference-based or entropy-based strategy, indicating that it is required to consider both the user preference and the capability of reducing candidate uncertainty when deciding the asked attribute. 
Without attribute selection, we observe that the impact on applications with small action space (e.g., LastFM and Yelp) is less than those with large action space (e.g., LastFM*, Yelp*, and E-Commerce). 
UNICORN is merely working without item selection, since there are no pretrained recommendation components in the framework and the preference-based item selection serves as an auxiliary item recall process. 

\begin{table}
\setlength{\abovecaptionskip}{2pt}   
\setlength{\belowcaptionskip}{2pt}
\centering
  \caption{The effect of hyper-parameters.}
\begin{tabular}{ccccccc}
\toprule 
Dataset&\multicolumn{3}{c}{LastFM*}&\multicolumn{3}{c}{E-Commerce}\\
\cmidrule(lr){2-4} \cmidrule(lr){5-7}
$K_v$&10&20&50&10&20&50\\
\midrule
$K_p$=1&0.333&0.321&0.306&0.202&0.194&0.186\\
$K_p$=10&\textbf{0.349}&0.305&0.297&\textbf{0.217}&0.205&0.189\\
$K_p$=20&0.298&0.286&0.254&0.202&0.199&0.178\\
\midrule

\cmidrule(lr){2-4} \cmidrule(lr){5-7}
$L_g$&1&2&3&1&2&3\\
\midrule
$L_s$=1&0.320&\textbf{0.349}&0.341&0.202&0.217&0.212\\
$L_s$=2&0.324&0.346&0.332&0.182&\textbf{0.221}&0.216\\
$L_s$=3&0.315&0.340&0.339&0.189&0.208&0.196\\
\bottomrule
\end{tabular}
\label{hyper}
\end{table}

\subsection{Parameter Sensitivity Analysis}
The upper part in Table~\ref{hyper} summarizes the experimental results  (hDCG) by varying the number of selected candidate actions. Since similar conclusions can also be drawn on other datasets, we only report the results on LastFM* and E-Commerce datasets due to space limit. As for the number of selected attributes, it is likely to discard the important attributes when only selected the attribute with the highest weighted entropy (e.g., $K_p$=1). However, within a certain training interaction period (10,000 episodes in our case), UNICORN generally achieves the best performance when only selecting a small number of candidate items for policy learning. The results also demonstrate the necessity of pruning the available actions when there is a large action search space in UCRPL.

The lower part shows the experimental results by varying the number of network layers. As for GCN, 1-hop aggregation ($L_g$=1) is not enough for capturing all the useful information for graph representational learning. However, there is no much difference between 2-hop ($L_g$=2) and 3-hop ($L_g$=3). Besides, for Transformer, we can see that 1-layer ($L_s$=1) is sufficient for a good performance.

\subsection{Qualitative Analysis}

In order to intuitively study the difference between the proposed UNICORN and other state-of-the-art CRS methods, we randomly sample a real-world interaction from the E-Commerce dataset. The generated conversations by UNICORN, SCPR, EAR, and CRM with the user simulator are presented in Figure~\ref{case}. Facing the large candidate action space, CRM tends to only trigger the recommendation component to make recommendations, and EAR continuously asks the questions that are not preferred by the user. Despite the success of SCPR in predicting user-preferred attributes, the policy learning in SCPR only decides when to ask or recommend, based on the number of candidate items, which leads to some unnecessary or redundant question-asking turns. UNICORN systematically addresses these issues by making a comprehensive decision of the next action. 

\section{Conclusions}
In this work, we formulate three separated decision-making processes in CRS, including when to ask or recommend, what to ask and which to recommend, as a unified policy learning problem.
To tackle the unified conversational recommendation policy learning problem, we propose a novel and adaptive RL framework, which is based on a dynamic weighted graph. In addition, we further design two simple yet effective action selection strategies to handle the sample efficiency issue.
Experimental results show that the proposed method significantly outperforms state-of-the-art CRS methods across four benchmark datasets and the real-world E-Commerce application with remarkable scalability and stability.

\bibliographystyle{ACM-Reference-Format}
\bibliography{sample-bibliography}

\end{document}